\documentclass{article}
\usepackage{epsfig}
\begin{document}

\title{\bf CP Violation and the Search for New Physics \footnote
{Invited talk at 3rd International Symposium on LHC Physics 
and Detectors (LHC 2001), Chia, Sardinia, Italy,
25-27 Oct 2001.}} 
\author{Gustavo C. Branco \footnote{e-mail: 
gbranco@alfa.ist.utl.pt }}

\maketitle

\begin{center}
Departamento de F\'\i sica 
Instituto Superior T\'{e}cnico, \\ Av. Rovisco Pais P-1049-001,
Lisboa, Portugal
\end{center}

\begin{abstract}
We present an overview of CP violation in the Standard Model and Beyond,
describing various possible sources of CP violation and how to search
for them.
\end{abstract}

\section{Introduction}
It is well known that CP violation can be introduced in the Standard Model
(SM) with three or more generations,
provided one allows for complex Yukawa couplings. For a long time the only 
experimental evidence for CP violation came from the Kaon sector. Recently,
one has had experimental evidence for CP violation also in the B sector.
Yet, CP violation continues being one of the least experimentally 
constrained aspects of the SM. Since the breaking of CP is closely 
related to one of the least theoretically understood sectors of the SM, 
namely the Higgs sector and 
the generation of fermion masses and mixings, it is clear that the study
of CP violation, both theoretically and experimentally, is likely to play 
a major r\^ ole in the future development of Particle Physics.

\subsection{Experimental evidence for CP violation}

The following three independent CP observables have been 
measured \cite{Groom:in}:

(i) Kaon sector:

\begin{eqnarray}
\vert \epsilon_k \vert = (2.28 \pm 0.02) \times 10^{- 3} \nonumber  \\
\epsilon'/\epsilon = (1.72 \pm 0.18) \times 10^{-3}
\end{eqnarray}

(ii) B sector
\begin{equation}
\sin(2 \beta ) = 0.75 \pm 0.09 \mathrm{(stat.)} \pm 0.04 \mathrm{(syst.)}
\; \; \mathrm{BaBar\; \cite{Aubert:2002gv}}
\end{equation}
\begin{equation}
\sin(2 \beta ) = 0.99 \pm 0.15 
\; \; \mathrm{Belle\; \cite{Abe:2001xe}}
\end{equation}

(iii) Baryon Asymmetry of the Universe (BAU)
\begin{equation}
{n_B}/{n_\gamma} = (1.5 - 6.3) \times 10^{- 10}
\end{equation}
The results (i), (ii) are in agreement with the Standard Model (SM) and its 
Kobayashi-Maskawa (KM) mechanism of CP violation. On the other hand, 
it is by now established that the strength of CP violation in the SM is not 
sufficient to generate the observed BAU \cite{bau}.

\subsection{The Strong CP problem}

An essential feature of 'tHooft's solution of the $U (1)$ 
problem \cite{'tHooft:1986nc},
is the fact that the QCD Lagrangian has to include a term:

\begin{equation}
{\cal L}_\theta = \theta_{QCD} \frac{g^{2}_{s}}{32 \pi ^2}
 F^{a \mu \nu} \tilde F^{a}_{\mu \nu} \label{cal}
\end{equation}
where $\tilde F^{\mu \nu} \equiv (1/2) \epsilon^{\mu \nu \rho \sigma} 
F_{\rho \sigma}$ and $\theta_{QCD}$ is a free parameter. 
The inclusion of this term is crucial for the solution of the $U (1)$ problem, 
but it leads to another difficulty, due to the fact that $\cal L _{\theta}$, 
violates P, T  and CP, while conserving C. In the SM, the quark mass matrices 
$M_{u}, M_{d}$ are generated through spontaneous symmetry breaking of the 
$SU (2) \times U (1)$ gauge symmetry and they are, in general, arbitrary 
complex matrices which are diagonalized by a bi-unitary 
transformation acting on 
left-handed and right-handed fields. These transformations include 
in particular 
the chiral transformation necessary to make $M_u, M_d$ diagonal, real 
which induces a contribution to $\theta_{QCD}$:
\begin{equation}
\theta_{QCD} \rightarrow \bar\theta \equiv \theta_{QCD} + \theta_{QFD}
\end{equation}
where
\begin{equation}
\theta_{QFD} = arg\ det (M_u M_d)
\end{equation}
As a result, $\bar\theta$ is the physical parameter which measures 
the strength of CP violation in non-perturbative QCD, leading in 
particular to an electric dipole moment (EDM) for the neutron. 
In chiral perturbation theory, the following result has been obtained 
\cite{Crewther:1979pi} for the neutron $EDM$\
\begin{equation}
D_n = (3.6 \times 10^{-16} \bar  \theta) \ e c m 
\end{equation}
The experimental bound on $D_n$ implies the following bound on $\bar \theta$:

\begin{equation}
\bar\theta < 10^{-10}
\end{equation}

Why should a dimensionless free parameter like $\bar \theta$, be so small? 
This is the so-called strong CP problem. Various solutions to the strong CP 
problem have been put forward, including the following:
\begin{itemize}
\item[(i)] $m_u = 0$ solution

This would be the simplest solution. If the up quark were exactly massless, 
then $\bar \theta$ could be set to zero by making a chiral transformation 
of the up quark field. The difficulty with this solution stems from the 
fact that it has been shown by Gasser and Leutwyler \cite{Gasser:1982ap}
that the up quark 
mass does not vanish. It is worth pointing out that a different point of 
view has been expressed by other authors \cite{kc} .

\item[(ii)] Peccei-Quinn solution
     
Peccei and Quinn have pointed out \cite{Peccei:1977hh}
that if there is a global chiral 
symmetry $U (1)_{PQ}$ under which both the quarks and the Higgs multiplets 
transform trivially, the $\bar \theta$ parameter becomes a dynamical 
variable which can be set to zero. It was pointed out by Weinberg and 
Wilczek \cite{ww}
that since $U (1)_{PQ}$ 
is a global continuous symmetry which is spontaneously broken by the vacuum, 
there is a Goldstone boson, named axion, which acquires mass through 
instanton effects. The original axion has been immediately ruled out
by experiment and so far, none of the its variants have been discovered.
The Peccei-Quinn suggestion is clearly one of the most elegant solutions 
to the strong CP problem. Its main drawback, is the fact that axions have
not been experimentally discovered yet. 

\item[(iii)] Solutions with calculable and naturally small $\bar \theta$

In this class of solutions, one imposes CP (or P) invariance at the 
Lagrangian level and chooses a Higgs potential so that CP (or P) is 
spontaneously broken. The CP (or P) invariance of the Lagrangian requires 
$\theta_{QCD} = 0$. If appropriate symmetries are added to the 
Lagrangian so that $\theta_{QFD}$ vanishes at tree level, 
then $\bar \theta$ equals zero at tree level and receives small, calculable 
contributions in higher orders. A specially attractive scenario in this 
class of theories is the one proposed by Barr and 
Nelson \cite{bn}, which can be 
implemented in the context of Grand Unified Theories or in the framework 
of $SU (2) \times U (1)$ \cite{Bento:ez} .
\end{itemize}
\begin{figure}[t]
\begin{center}
\epsfig{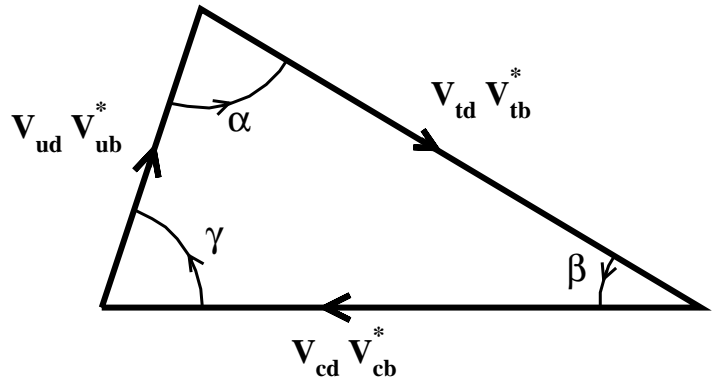}
\caption{The unitarity triangle in the complex plane} 
\end{center}
\end{figure}

\section{Motivation for New Physics contributions to CP Violation}

Although the SM and its KM mechanism of CP violation is in agreement with 
all the data in the Kaon and B sectors, there is motivation to consider 
New Physics contributions to CP violation, namely:

(i) The fact that the strength of CP violation in the SM is not 
sufficient to generate the observed BAU provides a strong motivation 
to consider new sources of CP violation, beyond the KM mechanism. 
Whether the new sources of CP violation needed to generate 
the correct BAU will be ``visible" at low energies 
(e.g. through the study of CP asymmetries in B-decays), 
is an open question. It depends very much on the origin of 
these new sources of CP violation. If these new contributions to 
CP breaking arise, for example, from low-energy supersymmetry or 
due to the presence of Z-mediated 
flavour-changing neutral currents \cite{Branco:1998yk},
then it is likely that they will manifest themselves as deviations 
from the SM predictions for CP asymmetries in B-decays. However one 
may generate the correct BAU through a completely different mechanism 
like, for example, baryogenesis through 
leptogenesis \cite{Fukugita:1986hr}. In this 
framework, a lepton asymmetry is initially generated by the decay 
of right-handed neutrinos and it is then converted into a baryon 
asymmetry by $(B + L)$ - violating sphaleron processes. In this 
case, the relevant sources of CP violation have to do with the 
coupling of right-handed neutrinos \cite{link}
and therefore the predictions 
of the SM for CP asymmetries in B-decays will not be affected.

(ii) Almost all extensions of the SM include new sources of CP 
violation. For example, in supersymmetric extensions of the SM, 
there is a large number of new phases which may play a r\^ole in 
CP asymmetries in the B system \cite{mvn} as well as through 
their contribution to various EDMS.

\section{CP violation in gauge theories}

There are two ways of introducing  CP violation in gauge 
theories \cite{Branco:1999fs}:

$\bullet$ {\bf At the Lagrangian level:}

In this case, the Lagragian is such that there is no transformation 
that may be physically interpreted as a CP transformation, under 
which the Lagrangian is invariant. The minimal model of this 
class is the 3 generation Standard Model with complex Yukawa couplings, 
where the Kobayashi-Maskawa (KM) mechanism takes place.

$\bullet$ {\bf Spontaneous CP violation:}

Two conditions have to be satisfied in order to 
achieve spontaneous CP violation:

(i) There is a transformation that may be physically interpreted as CP, 
under which the Lagrangian is invariant.

(ii) There is no transformation that may be physically interpreted 
as CP, under which both the Lagrangian and the vacuum are invariant.
 
The minimal model of this class is the Lee model, with two Higgs 
doublets and with flavour changing neutral currents
in the Higgs sector. We will discuss this model in subsection 3.2.

\subsection{CP violation in the Standard Model}

In the SM the Yukawa couplings are in general 
$n_g \times n_g$ complex matrices in flavour space, with $n_g$ denoting
the number of fermion generations.
Upon spontaneous symmetry breaking of the $SU (2) \times U (1)$ 
gauge symmetry, these couplings lead to complex quark mass matrices 
$M_u, M_d$ for the up and down quarks, respectively. 
These mass matrices are diagonalized by bi-unitary transformations:
\begin{eqnarray}
U_L^{u^{\dag}} M_u U_R^u = D_u \equiv diag. (m_u, m_c, m_t) \\
U_L^{d^{\dag}}M_d U_R^d = D_d \equiv diag. (m_d, m_s, m_b)
\end{eqnarray}
In the mass eigenstate basis, the charged currents can then be written as:
\begin{equation}
{\cal L}_W = {g \over \sqrt{2}} \bar u_L V_{CKM} \gamma_{\mu} 
d_L W^{\mu} + h. c.
\end{equation}
with $V_{CKM} = U_L^{u^{\dag}} U_L^d$. The mass terms are:
\begin{equation}
{\cal L}_{mass} = m_u \bar u u + m_c \bar c c + m_t \bar t t + 
m_d \bar d d + m_s \bar s s + m_b \bar b b.
\end{equation}
One has the freedom to rephase the quark fields:

\begin{eqnarray}
u_{\alpha} \longrightarrow u'_{\alpha} = e^{- i \varphi \alpha} 
u_{\alpha} \\
d_k \longrightarrow d'_k = e^{- i \psi k} d_k
\end{eqnarray}
Under the above rephasing, the mass terms remain invariant, 
while $V_{CKM}$ transforms as:
\begin{equation}
V_{\alpha k} \longrightarrow 
V'_{\alpha k} = e^{i (\psi_k - \varphi_{\alpha})} V_{\alpha k}
\end{equation}
where we have dropped the suffix $CKM$.

It is obvious that only rephasing invariant quantities have physical meaning. 
The simplest rephasing invariant functions of $V_{\alpha k}$ are moduli 
and invariant quartets:
\begin{equation}
\vert V_{\alpha k} \vert^2, \ \ \ Q_{\alpha i \beta j} 
\equiv V_{\alpha i} V_{\beta j} V_{\alpha j}^* V_{\beta i}^*
\end{equation}

Assuming a non-degenerate quark mass spectrum, the most general 
CP transformation for the quark fields which leaves the Lagrangian 
invariant, is:
\begin{eqnarray}
CP u_{\alpha} (t, \vec r) (CP)^\dagger  = e^{i \xi k} \gamma^0 C \bar 
u_{\alpha}^{T} (t, - \vec r)\\
CP d_k (t, \vec r) (CP)^\dagger  = e^{i \xi k} \gamma^0 C \bar 
d_k^T (t, - \vec r)
\end{eqnarray}

It can be readily verified that CP invariance constrains all rephasing 
invariant 
functions of $V$ to be real. In particular, CP invariance implies that 
$Im Q_{\alpha i \beta j} = 0$. For two generations there is only one 
quartet and unitarity constrains its imaginary part to vanish. Indeed
\begin{equation}
V_{u d} V_{cd}^* + V_{u s} V_{c s}^* = 0
\end{equation}
implies
\begin{equation}
V_{u s}^* V_{c s} V_{u d} V_{c d}^* + \vert V_{u s} 
\vert^2 \vert V_{c s} \vert^2 = 0
\end{equation}
and thus:
\begin{equation}
\mbox{Im} (V_{u d} V_{c s} V_{u s}^* V_{c d}^*) = 0
\end{equation}
For three or more generations, unitarity does not constrain 
$Im Q_{\alpha i \beta j}$ to vanish and therefore CP violation can arise.
This is the KM mechanism. Unitarity of $V_{CKM}$ plays a major r\^ole 
in restricting the strength of CP violation in the SM and it implies 
a series of relations which can be used to test the SM, with the potential 
of uncovering New Physics. Let us consider orthogonality of the first 
two rows of $V_{CKM}$:
\begin{equation}
V_{u d} V_{c d}^* + V_{u s} V_{c s}^* + V_{u b} V_{c  b}^* = 0
\label{vud} 
\end{equation}
Multiplying both sides of Eq. (\ref{vud}) by $V_{u s}^* V_{c s}$, 
and taking imaginary parts, one obtains:
\begin{equation}
Im (V_{u d} V_{c s} V_{u s}^* V_{c d}^*) = 
Im (V_{u s} V_{c b} V_{u b}^* V_{c s}^*)
\end{equation}
Similarly, one can prove that all $\vert Im Q \vert$ have the same 
value for any of the invariant quartets of the $3 \times 3$  $V_{CKM}$ matrix. 
This is a very special feature of the $3 \times 3$  unitary 
$V_{CKM}$ matrix, 
which would not hold true if, for example, there were four 
fermion generations. 
The quantity $\vert Im Q \vert$ can then be used as a measure of the 
strength of CP violation in the SM. Furthermore, $\vert Im Q \vert$ has a 
simple geometrical interpretation. To each of the six unitarity relations 
corresponding to orthogonality of rows and columns of  $V_{CKM}$, 
one can associate a 
triangle in the complex plane, as in Fig.1. 
All the unitarity triangles have the same area, 
which is proportional to $\vert Im Q \vert$:
\begin{equation}
Area = {1 \over 2} \vert Im Q \vert
\end{equation}
Under rephasing of quark fields, the triangles rotate and thus the 
orientation of the unitarity triangles has no physical meaning. 
However the internal angles of the unitarity triangles are invariant under 
rephasing and do have physical meaning. Of special interest, is the triangle 
corresponding to orthogonality of the first and third columns of 
the $CKM$ matrix, represented in Fig.1. This triangle has the special 
feature of having all sides of comparable size. Furthermore, 
in the context of the $SM$, the internal angles of this triangle are 
related to various $CP$ asymmetries in B-meson decays. 
The angles $\alpha, \beta, \gamma$ are represented in Fig.1 and are defined by:
\begin{eqnarray}
\alpha \equiv arg ( - {V_{td} V_{tb}^* \over V_{ud} V_{u b}^*} ) = 
arg (- Q_{ubtd}) \\
\beta \equiv arg ( - {V_{c d} V_{cb}^* \over V_{t d} V_{tb}^*} ) = 
arg (- Q_{tbcd}) \\
\gamma \equiv arg ( - {V_{ud} V_{ub}^* \over V_{cd} V_{c b}^*} ) = 
arg (- Q_{cbud}) \\
\end{eqnarray}
One can derive necessary conditions for CP invariance in the SM, with 
an arbitrary 
number of generations, written in terms of weak-basis (WB) invariants.

These invariants have the advantage that they can be evaluated in any WB. 
An example of such a WB invariant condition 
for CP invariance is \cite{Bernabeu:fc}:

\begin{equation}
tr [ H_u, H_d ]^3 = 0 \label{trh}
\end{equation}
where $H_u \equiv M_u M_u^{\dag}$, $H_d = M_d M_d^{\dag}$. For two fermion 
generations Eq. (\ref{trh}) is automatically satisfied with arbitrary 
Hermitian matrices $H_u, H_d$. This is, of course, to be expected since 
for the two generation SM, CP cannot be broken. For three generations, 
Eq. (\ref{trh}) 
is a necessary and sufficient condition for CP invariance.

This WB invariant can be evaluated in terms of quark masses 
and mixings:
\begin{equation}
tr [ H_u, H_d ]^3 = 6i (m_t^2 - m_c^2) (m_t^2 - m_u^2) (m_c^2 - m_u^2) 
(m_b^2 - m_s^2) (m_b^2 - m_d^2) (m_s^2 - m_d^2) Im Q \label{xyz} 
\end{equation} 
From Eq. (\ref{xyz}) it follows that in order for CP violation to take place, 
no two quarks of a given charge can be degenerate.
However, there is CP violation in the SM even in the limit where the 
mass of a given quark, for example $m_u$, vanishes.

\subsection{Spontaneous CP violation}

An alternative way of introducing CP, T violation is having a Lagrangian 
which is CP and T invariant but a vacuum which breaks CP, T. In the SM, 
it is not possible to achieve spontaneous CP violation (SCPV). The 
minimal extension of the SM where SCPV can be 
obtained is the Lee model \cite{Lee:iz}, 
where the gauge sector of the SM is not changed but in the Higgs sector, 
two scalar doublets are introduced. It has been shown that there is 
a range of parameters of the Higgs potential for which the minimum is at:
\begin{equation}
< 0 \vert \Phi_1 \vert 0 > = \left (
\begin{array}{l}
0 \\ v_1
\end{array}
\right ) ; \ \ \ < 0 \vert \Phi_2 \vert 0 > = \left (
\begin{array}{l}
0 \\ v_2 e^{i \theta}
\end{array}
\right ) \label{vev}
\end{equation}
The minimum of Eq. (\ref{vev}) does conserve electric 
charge but it violates 
CP and T. At the time when Lee suggested this model, there were only two 
fermion generations, so CP breaking originated only in Higgs mediated 
interactions. In the case of three fermion generations, it can be 
readily verified that the relative phase $\theta$ leads to CP 
violation both through neutral Higgs exchange and in charged weak 
interactions \cite{Branco:1999fs} .

This can be shown by considering the Yukawa interactions:
\begin{equation}
{\cal L}_Y = - \bar Q_L [ \Gamma_1 \Phi_1 + \Gamma_2 \Phi_2 ] 
d_R - \bar Q_L [ \Gamma_1^{'} 
\tilde {\Phi_1} + \Gamma_2^{'} \tilde {\Phi_2} ] u_R + h. c. 
\end{equation}
where $Q_L$ denote the left handed quark doublets. The Yukawa couplings  
$\Gamma_i, \Gamma_i^{'}$ are real matrices, 
so that CP invariance holds 
at the Lagrangian level. Upon spontaneous gauge symmetry breaking, 
one obtains the following quark mass matrices:
\begin{equation}
\begin{array}{ll}
M_d &= v_1 \Gamma_1 + v_2 \Gamma_2 e^{i \theta} \\
M_u &= v_1 \Gamma_1^{'} + v_2 \Gamma_2^{'} e^{- i \theta} \\ \label{ms}
\end{array}
\end{equation}
From Eqs. (\ref{ms}) one can compute $H_u, H_d$ and verify that the WB 
invariant of Eq. (\ref{xyz}) does not vanish in general, thus showing 
that in the Lee model with three fermion generations,
CP is violated by charged weak interactions through the KM 
mechanism, in spite of the existence a single phase in the model,
namely the phase $\theta$.
On the other hand, since quarks of a given charge receive 
mass from couplings to two different Higgs, there are flavour-changing 
neutral currents (FCNC) mediated by neutral Higgs. These couplings do 
lead to CP violation, as shown by Lee. In summary, in the Lee model
with three fermion generations and two Higggs doublets,
there are two sources of CP violation, the usual KM mechanism and
Higgs exchange.  The appearance of FCNC can be 
avoided by introducing extra discrete symmetries which constrain the 
Yukawa couplings in such a way that, for example, $\Phi_1$ only 
gives mass to down quarks while $\Phi_2$ only gives mass to 
the up quarks. In this case Higgs mediated neutral currents are 
naturally diagonal.
However, the introduction of this discrete symmetry
in the Lagrangian, forbids the presence of some terms in the 
Higgs potential in such a way that CP cannot be broken 
spontaneously \cite{Branco:1980sz}. 
One encounters a similar situation in  
the minimal supersymmetric standard model (MSSM), where CP cannot
be spontaneously broken. 
In the context of minimal extensions of the SM, one can have 
natural flavour conservation (NFC) in the Higgs sector 
(i.e., absence of FCNC due to a symmetry rather than by fine  
tuning) and yet achieve 
spontaneous CP violation \cite{Branco:1980sz}, 
by introducing a third Higgs 
doublet, which does not couple to quarks. In the context of 
supersymmetric extensions of the SM, spontaneous CP violation 
can be obtained in the next-to-minimal supersymmetric standard 
model (NMSSM)\cite{Branco:2000dq} with the introduction of 
a gauge singlet field.
In both the above described schemes, the imposition of NFC in 
the Higgs sector, 
together with the requirement of spontaneous CP violation, leads to a 
real CKM matrix \cite{Branco:1979pv}
and CP violation arises exclusively from physics 
beyond the SM. However, it should be emphasized that the above 
scenario of having a real CKM matrix is a very special case 
which results from the simultaneous requirement of spontaneous CP 
violation and NFC in the Higgs sector. When one considers extensions of 
the SM, the generic situation one encounters is having the coexistence 
of the KM mechanism with new sources of CP violation. As we have seen, 
this is the case of Lee's model, where besids the KM mechanism, one 
has a new source of CP violation, arising from Higgs exchange.

\section{The Search for New Physics}

Apart from its failure to account for the observed BAU, the SM and its 
KM mechanism of CP violation is in agreement with all the presently 
available experimental data. This agreement is an impressive success 
of the SM, which can be 
described in the following way. Let us adopt the standard 
parametrization of the CKM matrix, with three angles $\theta_i$ and one 
phase $\delta$ \cite{Groom:in}.
The three angles $\theta_1, \theta_2, \theta_3$ can be readily obtained 
from the knowledge of $\vert V_{us} \vert, \vert V_{ub} \vert, 
\vert V_{c b} \vert$ which can be extracted from Kaon and B-meson 
decays. Once these angles are fixed, one has to fit a large amount 
of data namely $\Delta M_{Bd}$, 
$\Delta M_{Bs}$, $\epsilon_k $, $\epsilon'/\epsilon$, 
and $a_ {J/\psi K_s}$, with a single parameter $\delta$. 
In the near future, 
BaBar and Belle will have much more precise 
data on  $a_ {J/\psi K_s}$ and 
hopefully will measure other CP asymmetries leading to the 
determination of $\gamma$. In LHC, the measurement 
of $\Delta M_{B_s}$ will be possible \cite{lhc}, 
as well as a significant improvement on the precision of all 
other asymmetries. When all this data will become available, the SM and 
in particular its KM mechanism of CP violation will be subjected to a 
very stringent test, with the potential for discovering New Physics. 
In considering the effects of New Physics, it is useful to make the 
following reasonable assumptions:

(i) One assumes that the quark decay amplitudes 
$\bar b \rightarrow  \bar c c \bar s$, 
$\bar b \rightarrow \bar u u \bar d$ 
as well 
as the semileptonic $b$ decays are dominated by the SM tree-level diagrams. 
This assumption is satisfied in most of the known extensions of the SM. 
In practice, this means that the extraction of $\vert V_{us} \vert, 
\vert V_{cb} \vert, \vert V_{ub} \vert$ from experimental data will 
continue being valid even in the presence of New Physics.

(ii) One allows for the possibility of having New Physics to 
$B_d - \bar {B_d}$ and $B_s - \bar {B_s}$ mixings. This is a 
reasonable assumption, because in the SM, $B - \bar B$ mixing 
receives contributions only at loop level and therefore New Physics 
can give additional contributions of comparable strength.

It is useful to parametrize the total contribution to the mixing as:
\begin{equation}
M^{(q)}_{12} =  [M^{(q)}_{12}]^{SM} r_q^2 e^{+ i \phi _q} 
\label{m12}
\end{equation}
where $q$ stands for $d, s$. From Eq. (\ref{m12}) it follows that 
$r_q \not = 1$ and or $\phi_q \not = 0$ signals the presence of 
New Physics. The main effect of the presence of New Physics contributions 
to $M^{(q)}_{12}$ is that the asymmetries $a_{J / \psi K_s}$, 
$a_{\pi + \pi -}$ 
will no longer measure the angles $\beta, \alpha$ but instead the following 
relation will apply:
\begin{eqnarray}
a_{J / \psi K_s} = \sin (2 \beta + \phi _q) \\
a_{\pi + \pi -} = \sin (2 \alpha - \phi _q)
\end{eqnarray}
Due to the constraints of unitary of the CKM matrix, the SM flavour 
sector is quite constrained since various measurable quantities 
(which are in general independent) are related within the framework of the SM. 
Any deviation from these SM relations will signal the presence of New Physics. 
An example of such a relation is the 
Aleksan-London-Kayser \cite{Aleksan:1994if} relation:
\begin{equation}
\sin \chi \cong {\vert V_{u s} \vert^2 \over \vert V_{u d} \vert^2} 
{\sin \beta \sin \gamma \over \sin (\gamma + \beta)} \label{akl}
\end{equation}
with $\chi$ defined by $\chi \equiv \arg ( - {V_{cb} V_{ts} 
V_{cs}^* V_{tb}^*} )$.
The importance of Eq. (\ref{akl}) has been emphasized by Silva and 
Wolfenstein \cite{Silva:1996ih}.

In a large class of models beyond the SM, in 
particular in the supersymmetric 
extension of the SM, the $3 \times 3$ unitarity of $V_{CKM}$  
continues to hold. 
However, it should be emphasized that unitarity of $V_{CKM}$ is an assumption 
which should 
be tested experimentally. Deviations of $3 \times 3$ 
unitarity naturally arise 
in models with vector-like isosinglet quarks \cite{vector}, 
i.e. quarks whose left-handed 
and right-handed components are both singlets under $SU (2) \times U (1)$.

\section{Conclusions} 

In the next few years crucial new data will be provided by the various 
B factories. With this new data, one will be able to test the flavour sector
of the SM to a great level of accuracy. These experimental tests 
have the potential of discovering New Physics, specially if one takes into
account that the SM is highly constrained. As we have emphasized, within 
the SM, a series of in principle independent measurable  quantities are
parametrized by a single parameter, 
namely the KM phase $\delta$. Even in the 
event that no deviations from the SM predictions are found, future data
on the various CP asymmetries as well as on $B_s -\bar B_{s}$ mixing 
will still have a great impact, since they will lead to a precise 
determination of the $V_{CKM}$ matrix, specially of
its smallest elements, namely $\vert V_{ub} \vert$, $\vert V_{td} \vert$.
This will in turn have important implications for theories of flavour. In 
some of these theories, family symmetries are introduced which lead to 
calculability of $V_{CKM}$ in terms of quark mass ratios. Some of 
these models differ from each other precisely in their predictions for
$\vert V_{ub} \vert$, $\vert V_{td} \vert$. Therefore,
the measurement of these matrix elements within the SM will have 
the potential to select the correct theory of flavour.
 
In conclusion, the future data from B-factories will
have a great impact on the physics of flavour and CP violation.

\section*{\bf Acknowledgments}

I am thankful to the organizers of the Workshop for the excellent program
and for the friendly atmosphere in the Meeting. Special thanks are due
to Carlo Bosio and Gino Saitta for the efforts to make the Workshop a
great success. I would like to thank Roger Cashmore for the kind invitation 
and for CERN support. Finally, I am thankful to M.N. Rebelo for suggestions
and a careful reading of the manuscript.

\end{document}